\documentclass[journal,onecolumn,draftclsnofoot]{IEEEtran}
\usepackage{cite,graphicx,subcaption,hyperref}
\usepackage{color}
\usepackage{placeins}
\usepackage{float}
\usepackage{tabularx,colortbl,booktabs}
\usepackage{amsmath,amsthm,amssymb,pifont}
\usepackage{dsfont}
\DeclareMathOperator*{\argmax}{arg\,max}

\begin{document}

\title{A JND-based Video Quality Assessment Model and Its Application}
\author{Haiqiang~Wang,~Xinfeng~Zhang,~Chao~Yang,~and~C.-C.~Jay~Kuo,~\IEEEmembership{Fellow,~IEEE}
\thanks{Haiqiang Wang, Xinfeng Zhang, Chao Yang and C.-C. Jay Kuo are with the Ming-Hsieh Department
of Electrical Engineering, Signal and Image Processing Institute, University of Southern California, Los Angeles, CA 90089 USA e-mail: \{haiqianw, xinfengz, yangchao\}@usc.edu, cckuo@sipi.usc.edu.}
}


\maketitle

\begin{abstract}

Based on the Just-Noticeable-Difference (JND) criterion, a subjective
video quality assessment (VQA) dataset, called the VideoSet, was
constructed recently.  In this work, we propose a JND-based VQA model
using a probabilistic framework to analyze and clean collected
subjective test data. While most traditional VQA models focus on
content variability, our proposed VQA model takes both subject and
content variabilities into account.  The model parameters used to
describe subject and content variabilities are jointly optimized by
solving a maximum likelihood estimation (MLE) problem.  As an
application, the new subjective VQA model is used to filter out
unreliable video quality scores collected in the VideoSet.  Experiments
are conducted to demonstrate the effectiveness of the proposed model.

\end{abstract}


\IEEEpeerreviewmaketitle

\section{Introduction}

Subjective quality evaluation is the ultimate means to measure quality
of experience (QoE) of users. Formal methods and guidelines for
subjective quality assessments are specified in various ITU
recommendations, such as ITU-T P.910 \cite{itu1999subjective}, ITU-R
BT.500 \cite{assembly2003methodology}, etc. Several datasets on video
quality assessment were proposed, such as the LIVE dataset
\cite{seshadrinathan2010study}, the Netflix Public dataset \cite{vmaf},
the VQEG HD3 dataset \cite{video2010report} and the VideoSet
\cite{wang2017videoset}. Furthermore, efforts have been made in
developing objective quality metrics such as VQM-VFD
\cite{wolf2011video}, MOVIE \cite{seshadrinathan2010motion} and VMAF
\cite{vmaf}.

Machine learning-based video quality assessment (VQA) systems rely
heavily on the quality of collected subjective scores. A typical
pipeline consists of three main steps. First, a group of human viewers
are recruited to grade video quality based on individual perception.
Second, noisy subjective data should be cleaned and combined to provide
an estimate of the actual video quality. Third, a machine learning
model was trained and tested on the calibrated datasets, and the performance was reported in terms
of evaluation criteria. They are called the data collection, cleaning
and analysis steps, respectively. In this work, we propose a novel
method for the data cleaning step, which is essential for a variety of video contents viewed by different
individuals. This is a challenging problem due to the following
variabilities.
\begin{itemize}
\item Inter-subject variability. Subjects may have a different vision
capability.
\item Intra-subject variability. The same subject may have different
scores against the same content in multiple rounds.
\item Content variability. Video contents have varying characteristics.
\end{itemize}

When each content is evaluated several times by different subjects, a
straightforward approach is to use the most common label as the true
label \cite{whitehill2009whose}. This problem was examined more
carefully in \cite{narwaria2018data}, which verified the distribution
assumptions required for parametric testing. Furthermore, it discussed
practical considerations and made recommendations in the testing
procedure. Based on the Just-Noticeable-Difference (JND) criterion, a
VQA dataset, called the VideoSet \cite{wang2017videoset}, was constructed
recently. Being motivated by \cite{li2017recover}, we develop a
probabilistic VQA model to quantify the influence of subject and content
factors on JND-based VQA scores. Furthermore, we show that our model is
more robust than the MOS-based model in \cite{li2017recover} for noisy
measurements cleaning.

The rest of this work is organized as follows. Related previous work is
reviewed in Sec.  \ref{sec:background}. The proposed JND-based VQA model
is introduced in Sec. \ref{sec:model}.  The parameters inference problem
is examined in Sec. \ref{sec:inference}.  Experimental results in data
cleaning are shown in Sec.  \ref{sec:results}.  Concluding remarks are
given in Sec.  \ref{sec:conclusion}.

\section{Review of Related Work}\label{sec:background}

The impacts of subject and content variabilities on video quality scores
are often analyzed separately. A $z$-score consistency test was used as
a preprocessing step to identify unreliable subjects in the VideoSet. Another method was proposed in
\cite{liu2012variational}, which built a probabilistic model for the
quality evaluation process and then estimated model parameters with a
standard inference approach. A subject model was proposed in
\cite{janowski2015accuracy} to study the influence of subjects on test
scores. An additive model was adopted and model parameters were
estimated using real data obtained by repetitive experiments on the same
content. More recently, a generative model was proposed in
\cite{li2017recover} that treated content and subject factors jointly by
solving a maximum likelihood estimation (MLE) problem. Their model was
developed targeting on the traditional mean-opinion-scores (MOS) data
acquisition process with continuous degradation category rating.

The JND-based VOA methodology provides a new framework for fine-grained
video quality scores acquisition. Several JND-based VQA datasets were
constructed \cite{wang2017videoset,jin2016jndhvei,mcl_jcv}, and JND
location prediction methods were examined in \cite{huang2017measure,
wang2017prediction}. Being inspired by \cite{li2017recover}, we develop
a JND-based VQA model that considers subject and content
variabilities jointly in this work. Then, we will show that this new
method provides a powerful data cleaning tool for JND-based VQA
datasets.

\section{Derivation of JND-based VQA Model}\label{sec:model}

Consider a VQA dataset containing $C$ video contents, where each source
video clip is denoted by $c$, $c=1, \cdots, C$. Each source clip is
encoded into a set of coded clips $d_{i}$, $i=0, 1, 2, \cdots, 51$,
where $i$ is the quantization parameter (QP) index used in the H.264/AVC
standard. By design, clip $d_{i}$ has a higher PSNR value than clip
$d_{j}$, if $i<j$, and $d_{0}$ is the losslessly coded copy of $c$. The
JND of this set of coded clips characterizes the distortion visibility
threshold with respect to a given anchor, $d_{i}$. Through subjective experiments, JND points can be obtained from a sequence of
consecutive noticeable/unnoticeable difference tests between clips pair
$(d_{i}, d_{j})$, where $ j \in \{i+1, \cdots, 51\}$.

\subsection{Binary Decisions in Subjective JND Tests}

The anchor, $d_{i}$, is fixed while searching for the JND location. With
a binary search procedure to update $d_{j}$, it takes at most $L = 6$ rounds to find the
JND location.  Here, we use $l$, $l = 1, \cdots, L$, to indicate the
round number and $s$, $s = 1, \cdots, S$, to indicate the subject index,
respectively. The test result obtained from subject $s$ at round $l$ on
content $c$ is a binary decision - noticeable or unnoticeable
differences. This is denoted by random variable $X_{c, s, l} \in \{0,
1\}$. If the decision is unnoticeable difference, we set $X_{c, s, l} =
1$.  Otherwise, $X_{c, s, l} = 0$. The probability of $X_{c, s, l}$ can
be written as
\begin{equation}\label{eq:confidence}
Pr(X_{c, s, l}=1) = p_{c, s, l} \mbox{ and } Pr(X_{c, s, l}=0) = 1-p_{c, s, l},
\end{equation}
where random variable $p_{c, s, l} \in [0,1]$ is used to model the
probability of making the ``unnoticeable difference'' decision at a given
comparison.

We say that a decision was made confidently if all subjects made the
same decision, no matter it was ``noticeable difference'' or
``unnoticeable difference''. One the other hand, a decision was made
least confidently if two decisions had equal sample size. In light
of these observations, $p_{c, s, l}$ should be closer to zero for smaller
$l$ since the quality difference between two clips is more obvious in
earlier test rounds. It is close to 0.5 for larger $l$ as the coded clip
approaches the final JND location.

\subsection{JND Localization by Integrating Multiple Binary Decisions}

During the subjective test, a JND sample was obtained through multiple
binary decisions. Let $\mathbf{X}_{c,s}=[X_{c,s,1}, \cdots, X_{c,s,L}]$ denote
a sequence of decisions made by subject $s$ on content $c$. Random
variable $X_{c, s, l}$ is assumed to be independently identically
distributed (i.i.d) in subject index $s$. Furthermore, $X_{c, s, l}$ is
independent of content index $c$ since the binary search approaches the
ultimate JND location at the same rate regardless of the content. The
search interval at round $l$, denoted by $\Delta QP_l$, can be expressed as
\begin{equation}\label{eq:interval}
\Delta QP_l= \Delta QP_0 (\frac{1}{2})^l,
\end{equation}
where $\Delta QP_0=51$ is the initial JND interval for the first JND. The anchor location is $QP_{0}$, i.e. the reference and the JND is searched between $[QP_{1}, QP_{51}]$. We skip
comparison between clips pair $(QP_{0}, QP_{51})$ since it is a
trivial one.

By definition, the JND location is the coded clip that is the transition
point from the unnoticeable difference to the noticeable difference
against the anchor.  It is located at the last round after a sequence of
``noticeable difference'' decisions. Thus, the JND location on content
$c$ for subject $s$ can be obtained by integrating searching intervals
based on decision sequences $\mathbf{X}_{c,s}$ as
\begin{equation} \label{eq:summation}
Y_{c, s} = \sum_{l=1}^{L} X_{c, s, l} \Delta QP_l,
\end{equation}
since we need to add $\Delta QP_l$ to the offset (or the left ending)
point of the current searching interval if $X_{c, s, l}=1$ and keep the
same offset if $X_{c, s, l}=0$. The distance between the left end
point of the search interval and the JND location is no larger than one
QP when the search procedure converges. Then, the JND location could be
expressed as a function of the confidence of subject $s$ on content $c$:
\begin{equation} \label{eq:decompose}
Y_{c, s} = \Delta QP_0 \sum_{l=1}^{L} p_{c, s, l} (\frac{1}{2})^l.
\end{equation}

\subsection{Decomposing JND into Content and Subject Factors}

The JND locations depend on several causal factors: 1) the bias of the
subject, 2) the consistency of a subject, 3) the averaged JND location,
4) the difficulty of a content to evaluate. To provide a closed-form
expression of the JND location, we adopt the following probabilistic
model for the confidence random variable:
\begin{equation} \label{eq:factor}
p_{c, s, l} = \mu_{l} + \alpha\epsilon_{c} + \beta\delta_{s},
\end{equation}
where $\mu_{l} = \frac{1}{2}(1+e^{-\gamma l})$ is the averaged
confidence, $\epsilon_{c} \sim \mathcal{N}(\mu_{c}, \sigma_{c}^{2})$ and
$\delta_{s} \sim \mathcal{N}(\mu_{s}, \sigma_{s}^{2})$ are two Gaussian
random variables to capture content and subject factors, respectively.
$\alpha$ and $\beta$ are weights to control the effects of mentioned
factors. We set $\gamma=0.7$, $\alpha=1$ and $\beta=1$ empirically.

By plugging Eq. (\ref{eq:factor}) into Eq.
(\ref{eq:decompose}), we can express the JND location as
\begin{equation} \label{eq:final_decom}
Y_{c, s} = y_{c} + \mathcal{N}(0, v_{c}^{2}) + \mathcal{N}(b_{s}, v_{s}^{2})
\end{equation}
where $y_{c} = \Delta QP_{0} \sum_{l=1}^{L} (\frac{1}{2})^l
(\mu_l+\mu_{c})$ and $v_{c}^{2} = \kappa^{2}\sigma_{c}^{2}$ are content
factors. $b_{s} = \kappa\mu_{s}$ and $v_{s}^{2} = \kappa^{2}
\sigma_{s}^{2}$ are subject factors. $\kappa = \Delta QP_{0} \sum_{l=1}^{L} (\frac{1}{2})^{l} \approx 50$ is a constant.

The JND-based VQA model in Eq. (\ref{eq:final_decom}) decomposes the JND
location into the content term and the subject term.  The difficulty of
a content is modeled by $v_{c}^{2} \in [0, \infty)$. A larger
$v_{c}^{2}$ value means that its masking effect is stronger and the most
experienced experts still have difficulty in spotting artifacts in
compressed clips. The bias of a subject is modeled by parameter $b_{s}
\in(-\infty, +\infty)$. If $b_{s}<0$, the subject is more sensitive to
quality degradation in compressed video clips.  If $b_{s}>0$, the
subject is less sensitive to distortions. The sensitivity of an averaged
subject has a bias around $b_{s}=0$. Moreover, the subject variance,
$v_{s}^{2}$, captures the consistency of subject $s$. A consistent
subject evaluates all sequences deliberately.

\begin{figure*}[!h]
\captionsetup[subfigure]{labelformat=empty}
\centering
  \begin{subfigure}[b]{0.15\linewidth}\includegraphics[width=1.0\linewidth]{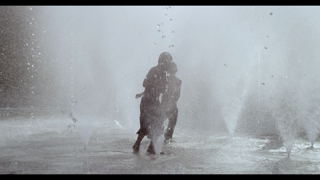}\caption{011}\end{subfigure}
  \begin{subfigure}[b]{0.15\linewidth}\includegraphics[width=1.0\linewidth]{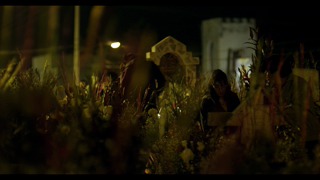}\caption{026}\end{subfigure}
  \begin{subfigure}[b]{0.15\linewidth}\includegraphics[width=1.0\linewidth]{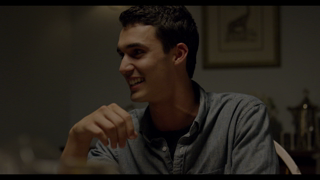}\caption{041}\end{subfigure}
  \begin{subfigure}[b]{0.15\linewidth}\includegraphics[width=1.0\linewidth]{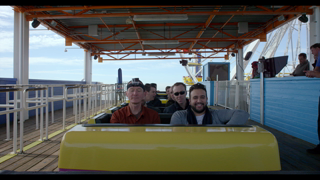}\caption{056}\end{subfigure}
  \begin{subfigure}[b]{0.15\linewidth}\includegraphics[width=1.0\linewidth]{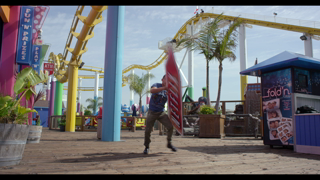}\caption{071}\end{subfigure}\\
  \begin{subfigure}[b]{0.15\linewidth}\includegraphics[width=1.0\linewidth]{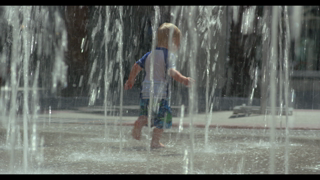}\caption{086}\end{subfigure}
  \begin{subfigure}[b]{0.15\linewidth}\includegraphics[width=1.0\linewidth]{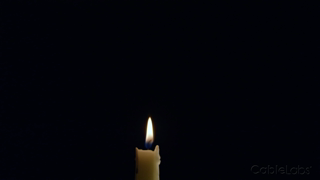}\caption{101}\end{subfigure}
  \begin{subfigure}[b]{0.15\linewidth}\includegraphics[width=1.0\linewidth]{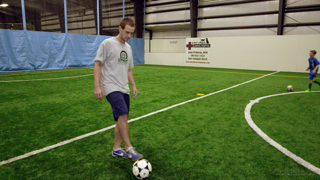}\caption{116}\end{subfigure}
  \begin{subfigure}[b]{0.15\linewidth}\includegraphics[width=1.0\linewidth]{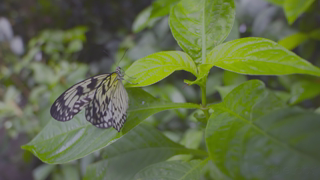}\caption{131}\end{subfigure}
  \begin{subfigure}[b]{0.15\linewidth}\includegraphics[width=1.0\linewidth]{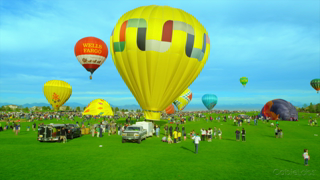}\caption{146}\end{subfigure}\\
  \begin{subfigure}[b]{0.15\linewidth}\includegraphics[width=1.0\linewidth]{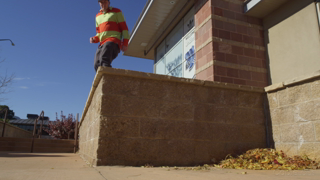}\caption{161}\end{subfigure}
  \begin{subfigure}[b]{0.15\linewidth}\includegraphics[width=1.0\linewidth]{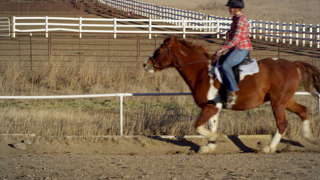}\caption{175}\end{subfigure}
  \begin{subfigure}[b]{0.15\linewidth}\includegraphics[width=1.0\linewidth]{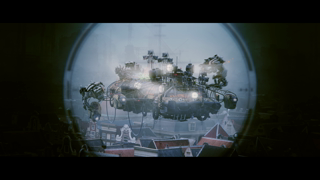}\caption{189}\end{subfigure}
  \begin{subfigure}[b]{0.15\linewidth}\includegraphics[width=1.0\linewidth]{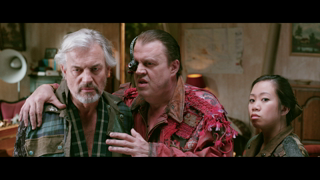}\caption{203}\end{subfigure}
  \begin{subfigure}[b]{0.15\linewidth}\includegraphics[width=1.0\linewidth]{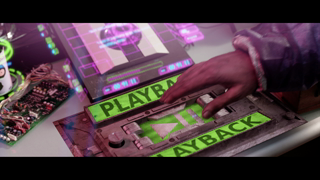}\caption{217}\end{subfigure}
\caption{Representative frames from 15 source contents.} \label{fig:thumbnails}
\end{figure*}

\section{Parameter Estimation}\label{sec:inference}

The JND-based VQA model in Eq. (\ref{eq:final_decom}) has a set of
parameters to determine; namely, $\theta=(\{y_c\}, \{v_{c}\}, \{b_s\},
\{v_{s}\})$ with $c = 1, \cdots, C$ and $s = 1, \cdots, S$. Under the
assumption that contents and subjects are independent factors on
perceived video quality, the JND location can be expressed by the
Gaussian distribution in form of
\begin{equation}\label{eq:jnd_gaussian}
Y_{c, s} \sim \mathcal{N} (\mu_{c,s}, \sigma_{c,s}^{2}),
\end{equation}
where $\mu_{c,s}=y_{c}+b_{s}$ and $\sigma_{c,s}^{2}=v_{c}^{2}+
v_{s}^{2}$.  The task is to estimate unknown parameters jointly given
observations on a set of contents from a group of subjects. A standard
inference method to recover the true MOS score was studied in
\cite{li2017recover}. Here, we extend the procedure to estimate the
parameters in the JND-based VQA model.

Let $L(\theta)=\log p(\{y_{c,s}\}|\theta)$ be the log-likelihood
function. One can show that the optimal estimator of $\theta$ is given
by $\hat{\theta}=\argmax_{\theta} L(\theta)$. By omitting constant terms, we
can express the log-likelihood function as
\begin{align}
  \begin{split}
    L(\theta) &= \log p(\{y_{c,s}|\theta)\\
              &= \log \prod_{c,s} p(y_{c,s}|y_c, b_s, v_c, v_s)\\
              &= \sum_{c,s} \log p(y_{c,s}|y_c, b_s, v_c, v_s)\\
              & \equiv - \log(v_{c}^{2}+v_{s}^{2}) - \frac{(y_{c,s}-y_c-b_s)^2}{v_{c}^{2}+v_{s}^{2}}.
  \end{split}
\end{align}
The first and second order derivatives of $L(\theta)$ with respect to
each parameter can be derived. They are used to update the parameters
at each iteration according to the Newton-Raphson rule, i.e. $\theta
\leftarrow \theta - \frac{\partial L/\partial \theta}{\partial^{2}
L/\partial \theta^{2}}$.

\section{Experimental Results}\label{sec:results}

In this section, we evaluate the performance of the proposed model using
real JND data from the VideoSet and compare it
with another commonly used method. For reproducibility, the source code
of the proposed model is available at:
\url{https://github.com/JohnhqWang/sureal}.

\begin{figure*}[!h]
\captionsetup[subfigure]{labelformat=empty}
\centering
  \begin{subfigure}[b]{0.32\linewidth}
	   \centering{}\includegraphics[width=1.0\linewidth]{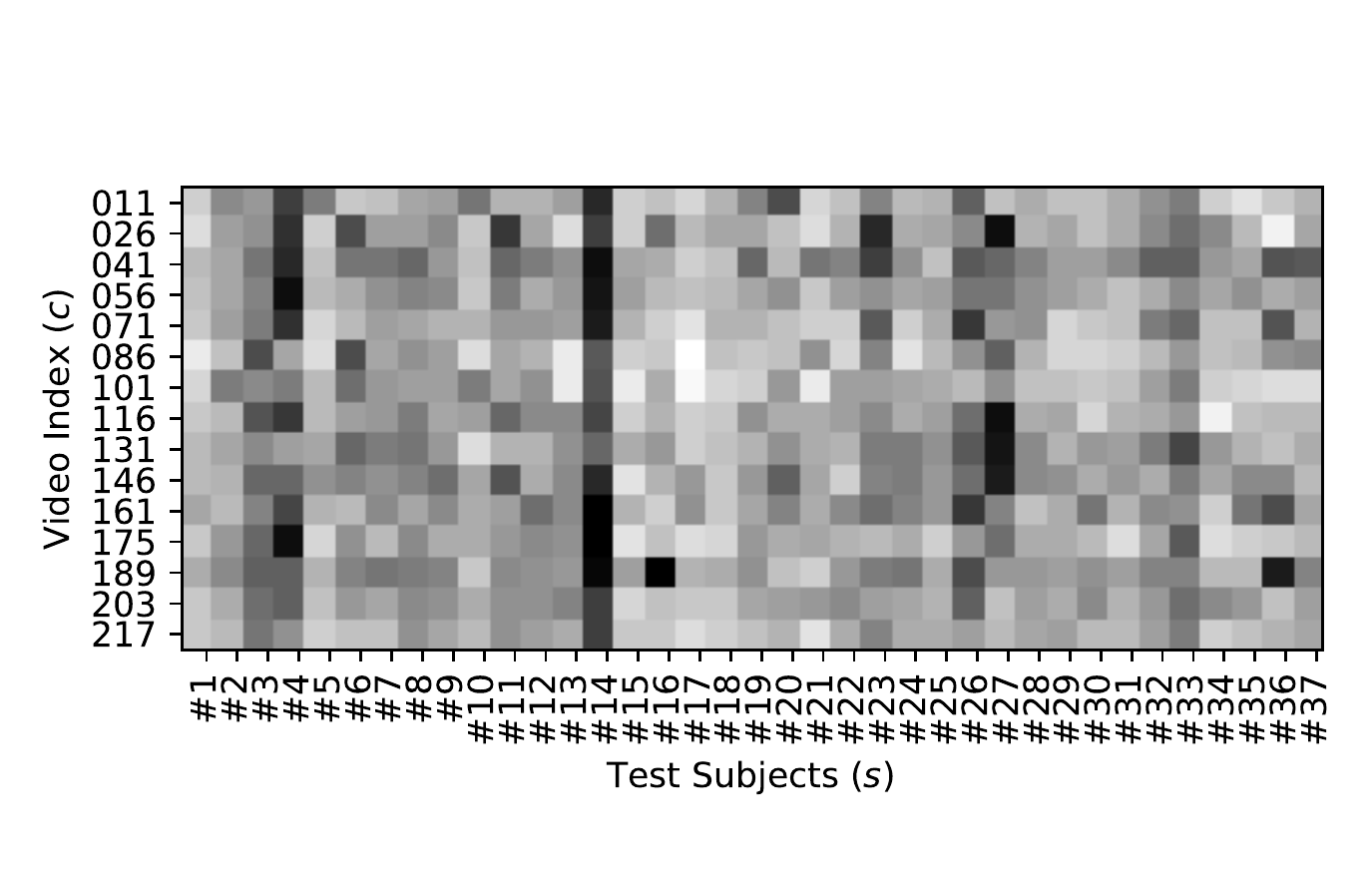}
     \subcaption{(a)}\label{fig:noisy_ycs}
  \end{subfigure}
  \begin{subfigure}[b]{0.3\linewidth}
	   \includegraphics[width=1.0\linewidth]{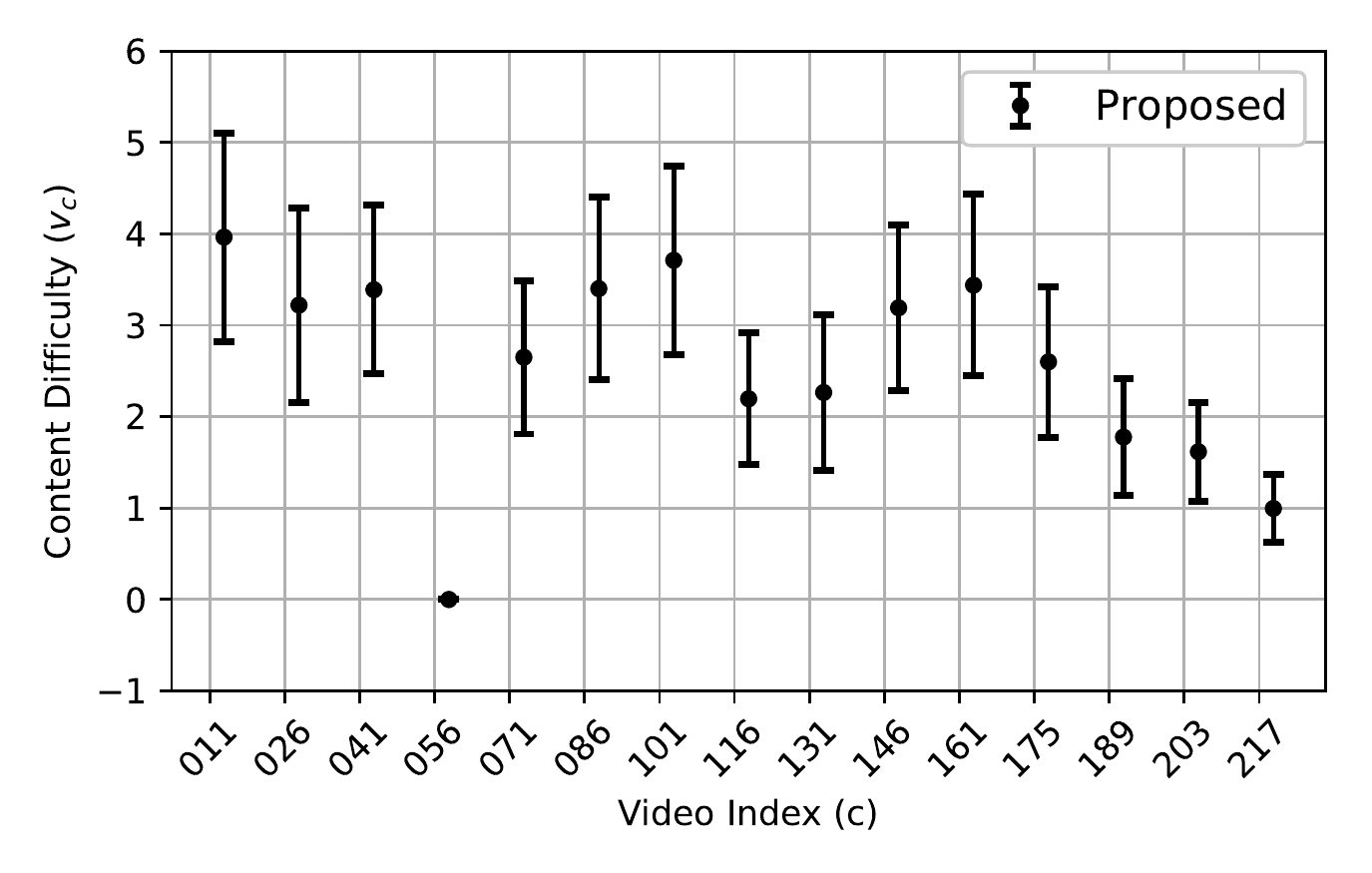}
     \subcaption{(c)}\label{fig:noisy_content_difficulty}
	\end{subfigure}
  \begin{subfigure}[b]{0.3\linewidth}
	   \centering{}\includegraphics[width=1.0\linewidth]{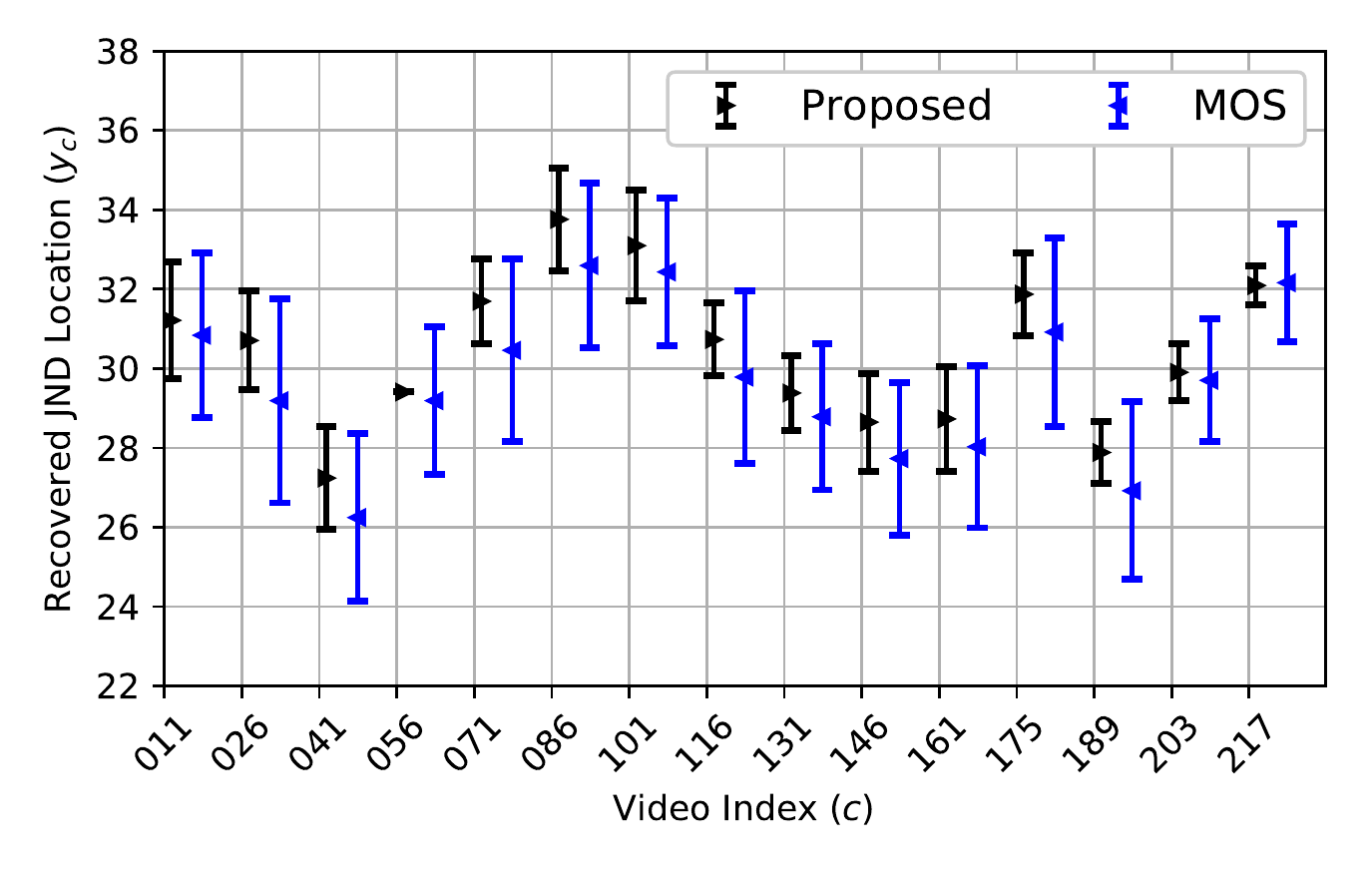}
     \subcaption{(e)}\label{fig:noisy_yc_mle_mos}
	\end{subfigure}
  \begin{subfigure}[b]{0.3\linewidth}
	   \centering{}\includegraphics[width=1.0\linewidth]{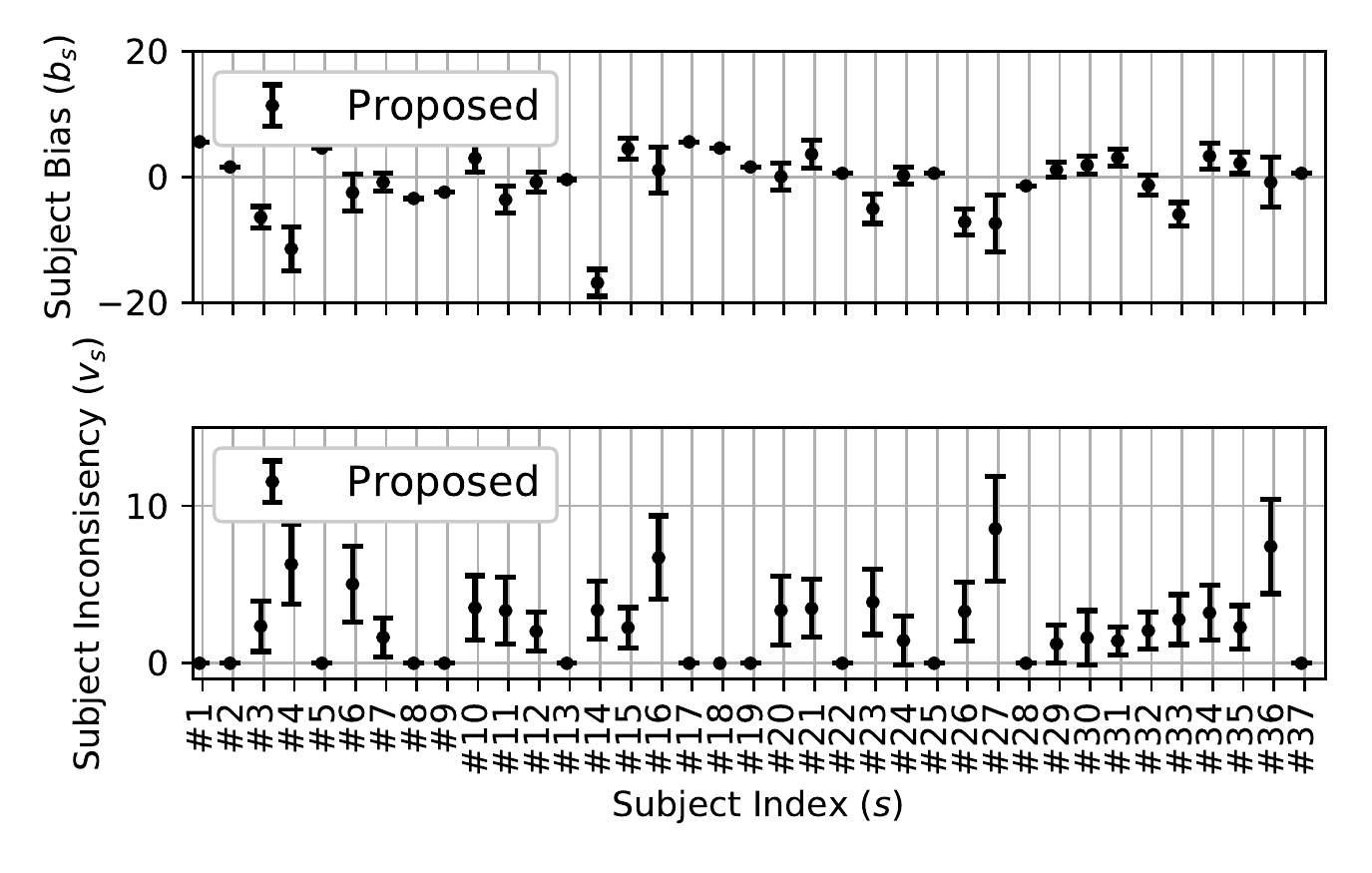}
     \subcaption{(b)}\label{fig:noisy_subj}
	\end{subfigure}
  \begin{subfigure}[b]{0.3\linewidth}
	   \centering{}\includegraphics[width=1.0\linewidth]{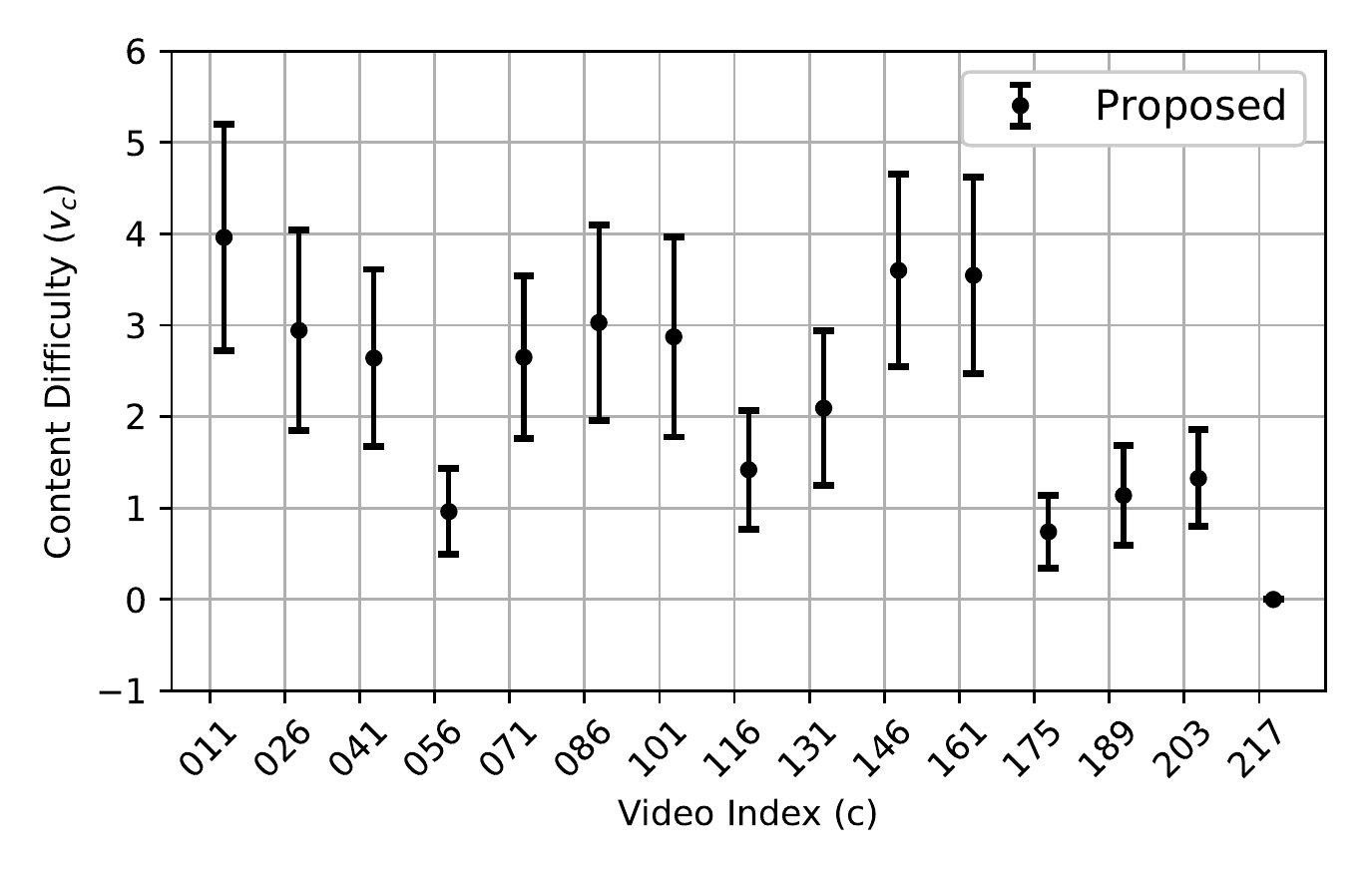}
     \subcaption{(d)}\label{fig:clean_content_difficulty}
	\end{subfigure}
  \begin{subfigure}[b]{0.3\linewidth}
	   \centering{}\includegraphics[width=1.0\linewidth]{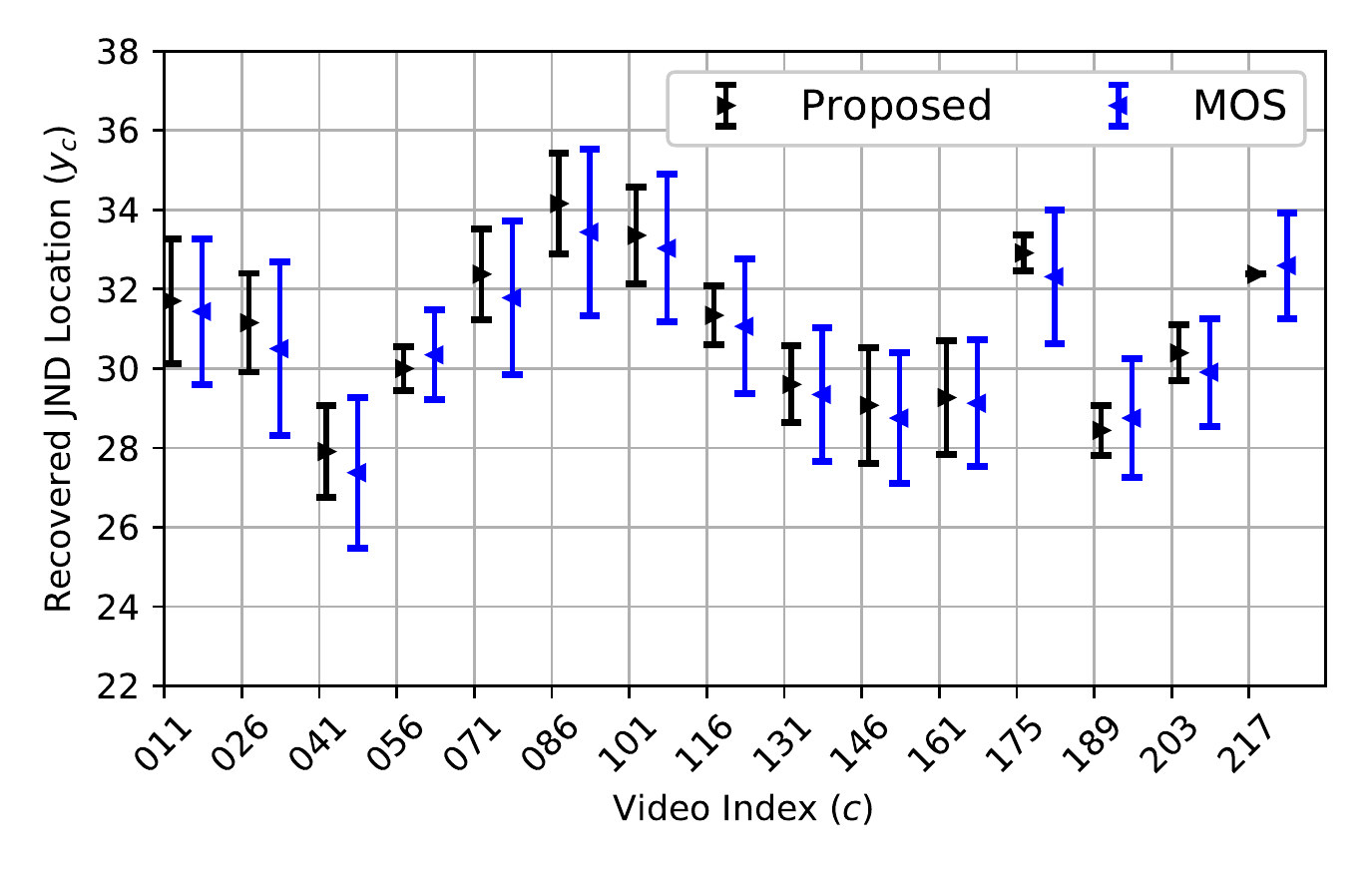}
     \subcaption{(f)}\label{fig:clean_yc_mle_mos}
	\end{subfigure}
\caption{Experimental results: (a) raw JND data, where each pixel
represents one JND location and a brighter pixel means the JND occurs
at a larger QP; (b) estimated subject bias and inconsistency on raw JND
data; (c) and (d) estimated content difficulty based on raw and cleaned
JND data, respectively, using the proposed VQA+MLE method; (e) and (f)
estimated JND locations based on raw and cleaned JND data, respectively,
using both the proposed VQA+MLE method and the MOS method. Error bars in
all subfigures represent the $95\%$ confidence
interval.}\label{fig:noisy_subject_factor}
\end{figure*}

\subsection{Experiment Settings}

The VideoSet contains 220 video contents in 4 resolutions and three JND
points per resolution per content. During the subjective test, the
dataset was split into 15 subsets and each subset was evaluated
independently by a group of subjects. We adopt a subset of the first JND
points on 720p video in the experiment. The subset contains 15 video
contents evaluated by 37 subjects. One representative frame from each of
the 15 video clips is shown in Fig. 1.  The measured
raw JND scores are shown in Fig. 2(a).

Standard procedures have been provided by the ITU for subject screening
and data modeling. For example, a subject rejection method was proposed
in the ITU-R BT.500 Recommendation \cite{assembly2003methodology}. The
differential MOS was defined in the ITU-T P.910 Recommendation
\cite{itu1999subjective} to alleviate the influence of subject and
content factors. However, these procedures do not directly apply to the
collected JND VQA data due to a different methodology. Traditional
VQA subjective tests evaluate video quality by a score while the JND-based
VQA subjective tests target at the distortion visibility threshold.

Here, we compare the proposed VQA model whose parameters are estimated
by the MLE method with those estimated by the standard MOS approach
\cite{li2017recover} in two different settings. First, we compare them
based on raw JND data without any cleaning process. Second, we clean unreliable data using the proposed VQA model and compare these two models with the cleaned JND data.

\subsection{Experiments on Raw JND Data}

The first experiment was conducted on the raw JND data without outlier
removal. Some subjects completed the subjective test hastily without
sufficient attention. By jointly estimating the content and subject
factors, a good VQA data model can identify such outlying quality
ratings from unreliable subjects.  The estimated subject bias and
inconsistency are shown in Fig. 2(b). The proposed JND-based VQA model
indicates that the bias of subjects \#04 and \#16 is very significant
(more than 10 QPs) as compared with others.  Furthermore, the proposed
model suggests that subjects \#16, \#26 and \#36 exhibit large
inconsistency. The observation was evidenced by the noticeable dark dots
on some contents.

Fig. 2(c) shows the estimated content difficulty. Content \#11 is a
scene about toddlers playing in a fountain. The masking effect is strong
due to water drops in the background and moving objects. Thus,
compression artifacts are difficult to perceive and it has the highest
content difficulty. On the other hand, content \#203 is a scene
captured with still camera. It focuses on speakers with still blurred
background. The content difficulty is low as the masking effect is weak
and compression artifacts are more noticeable.

The estimated JND locations using the proposed VQA+MLE method and the
MOS method are compared in Fig. 2(e). The proposed method offers more
confident estimates as its confidence intervals are much tighter than those estimated by the MOS method. More importantly, the estimates
by the proposed method are significantly different from those by the MOS
method. It is well known that the mean value is vulnerable to outliers
for a fixed sample size. The proposed method is more robust to noisy
subjective scores, which tend to have a negative bias in general.

\subsection{Experiments on Cleaned JND Data}

Here, we remove the outlying JND samples detected by the proposed model.
They are from subjects with a larger bias value or inconsistent
measures.  We show the estimated content difficulty in Fig. 2(d) and
compare the estimated JND locations of the proposed method and the MOS
method in Fig. 2(f) on the cleaned dataset.  We see that the proposed
VQA+MLE method can estimate the relative content difficulty accurately.
We also notice that the estimation changed a lot for some contents. The
reason is that considerable portion ($5/37=13.5\%$) of the subjects were
removed. The bias and inconsistency of the removed scores have great
influence on the conclusion of these contents.

By comparing Figs. 2(e) and 2(f), we observe that outlying
samples changed the distribution of recovered JND locations in both
methods.  First, the confidence intervals of the MOS method decrease a
lot. It reflects the vulnerability of the MOS method due to noisy
samples.  In contrast, the proposed VQA+MLE method is more robust.
Second, the recovered JND location increases by 0.5 to 1 QP in both
methods after removing noisy samples. It demonstrates the importance of
building a good VQA model and using it to filter out noisy samples.

\section{Conclusion and future work}\label{sec:conclusion}

A JND-based VQA model was proposed to analyze measured JND-based VQA
data. The model considered subject and content variabilities, and
determined its parameters by solving an MLE problem iteratively. The
technique can be used to remove biased and inconsistent samples and
estimate the content difficulty and JND locations. It was shown by
experimental results that the proposed methodology is more robust to
noisy subjects than the traditional MOS method.

The MLE optimization problem may have multiple local maxima and the
iterative optimization procedure may not converge to the global maximum.
We would like to investigate this problem deeper in the future. Also, we
may look for other parameter estimation methods that are more efficient
and robust.

\bibliographystyle{IEEEtran}
\bibliography{refs}

\end{document}